\documentclass{article}
\usepackage{graphicx}
\usepackage{amsmath}
\usepackage{amsbsy}
\usepackage{epstopdf, epsfig}
\usepackage[a4paper, total={6in, 8in}]{geometry}
\usepackage{xspace}
\newcommand{\co}{CO\ensuremath{_\textrm{2}}\xspace}
\newcommand{\pcs}{\ensuremath{P_\textrm{c}}-\ensuremath{ S}\xspace}

\usepackage[colorlinks  = true,
            linkcolor   = blue,
            urlcolor    = blue,
            citecolor   = red,
            anchorcolor = blue]{hyperref}
\usepackage{ marvosym }   
\usepackage{amssymb}
\usepackage{subfig}
\usepackage{booktab}
\usepackage{hyperref}
\usepackage{amsthm}
\newtheorem{theorem}{Theorem}
\newtheorem{remark}{Remark}

\title{Implicit linearization method for non-standard two-phase flow in porous media}

\author{Abay M. Kassa$^{1,2}$ \and
  K. Kumar$^{1}$ \and
  Sarah E. Gasda$^{2}$ \and  
 \and F. A. Radu$^{1}$}
\date{}
\begin{document}
\maketitle
\noindent ${}^1$ Department of Mathematics, University of Bergen, P. O. Box 7800, 5020 Bergen, Norway.\\[5pt]
${}^2$ NORCE, Nyg{\aa}rdsgaten 112, 5008 Bergen, Norway.\\[5pt]
Corresponding author: Abay M. Kassa (E-mail: abka@norceresearch.no)
\begin{abstract}
In this paper, we consider a non-local (in time) two-phase flow model. The non-locality is introduced through  the wettability alteration induced dynamic capillary pressure function. We present a monotone fixed-point iterative linearization scheme for the resulting non-standard model. The scheme treats the dynamic capillary pressure functions semi-implicitly and introduces an $L$-scheme type \cite{List2016,Radu2015} stabilization term in the pressure as well as the transport equations.  We prove the convergence of the proposed scheme theoretically under physically acceptable assumptions, and verify the theoretical analysis with numerical simulations.  The scheme is implemented and tested for a variety of reservoir heterogeneity in addition to the dynamic change of the capillary pressure function. The proposed scheme satisfies the predefined stopping criterion within few number of iterations. We also compared the performance of the proposed scheme against the iterative IMplicit Pressure Explicit Saturation  scheme
\end{abstract}

\section{Introduction}
Unsaturated groundwater flow, enhanced oil recovery,   and subsurface carbon-dioxide (\co) storage \cite{Bear1998, Chen2006, Dagan1991, David2016, Nielsen, Nordboton2012} are typical applications of multi-phase porous media flow with high societal relevance. Numerical simulations including mathematical modeling and numerical methods have been applied to understand such flow processes. The governing mathematical models  are  highly non-linear and possibly degenerate systems of partial differential equations. Usually, the non-linearities are introduced through constitutive models such as relative permeabilities--and capillary pressure--saturation relations. We describe these relations by either van Genuchten \cite{RefvanGenuchten} or Brooks-Corey \cite{RefBrooksCorry, Dagan1991} parametrizations.  These parametrizations are only suited for rock surfaces that  experience static and uniform wetting property.  

In this paper, we focus on two-phase flow that considers dynamic pore-scale wettability alteration (WA) processes. 
WA mechanisms have been exploited in the petroleum industry, where optimal wetting conditions in the reservoir are obtained through a variety of means that include chemical treatment, foams, surfactants and low-salinity water flooding \cite{Morrow75, Morrow86, Buckley1998, Jabhunandan95, RefHaag17, RefSingh16}.  The WA processes are assumed to be instantaneous in the above studies.  Here, rather, we considered  exposure time-dependent WA mechanisms. In our previous work  \cite{Kassa2018}, we upscaled time-dependent WA processes to Darcy-scale phenomenon, and we have developed an interpolation based dynamic capillary pressure model. The proposed model is macroscale fluid history and time dependent (see section \ref{relacap} and \cite{Kassa2018} for the   details) in addition to the current wetting phase saturation. This implies that a non-local
capillary diffusion term in time is introduced in a two-phase flow model. These all impose an additional complexity onto the standard two-phase porous media flow model.  

Due to the non-linearity and dynamic heterogeneity of the designed model, it is impossible to derive analytic solutions. As a consequence, numerical approach is the only option to predict such flow dynamics. However, developing efficient algorithms for finding numerical solutions is also  a challenge in itself even for standard models \cite{Larisa2008}.  Besides the non-linearity and heterogeneity of the designed model, long-term temporal dynamics adds an extra difficulty for proposing a reliable numerical model. Implicit discretization in time has been employed to handle long-term subsurface evolution as it allows large time step size. Newton-type methods are usually applied to solve the resulting non-linear system of equations. These approaches are  second-order convergent (if converges). However, this order of convergence comes at a price of a costly computation of Jacobian of a system  at each time step \cite{Monteagudo2007, Chen2004, Anna, Radu2015}. Additionally, these methods are  only locally convergent \cite{Radu2015,Kou2010}.

The other alternative approach is the splitting and then coupling (splitting-coupling) scheme. It splits the entire system into sub-systems. The decomposed sub-problems are then solved sequentially and are coupled by data exchanges at each time step. The IMplicit Pressure Explicit Saturation (IMPES) scheme is a widely used splitting-coupling approach to model two-phase flow and component transport processes \cite{Anna, Radu2015, Chen2004, Nordboton2012, Kou2010b,Kou2010, Coats2003}. IMPES solves the pressure equation implicitly and updates the saturation explicitly. This approach eliminates the non-linear terms in the pressure and saturation equations by evaluating them at saturation and fluid properties up-winded from previous time step. As a consequence, the scheme is conditionally stable, and hence it requires sufficiently  small time step size to approximate the  solution. 

Several techniques can be implemented to improve the IMPES approach. 
A very straightforward approach imposes large time step for the pressure and then subdivides the time step size for the transport equation \cite{Chen2004,Coats2003}. This approach relies on the assumption that the reservoir  pressure changes slowly in time  compared to saturation evolution.  
The other approach solves the transport equation implicitly using Newton  method while the pressure is treated in the same way as the classical IMPES \cite{Tseng2000, WHEELER2003}. In Kou et al., \cite{Kou2010}, the capillary pressure function in the pressure equation is approximated by a linear function. This helps to couple the pressure and saturation equations at the current time step. However, the scheme involves calculation of matrix inverse and multiple number of matrix multiplications, which greatly increases the computational cost of the scheme. Furthermore, the transport equation is still solved explicitly in time, and the scheme is reduced to the classical IMPES when the capillary pressure is neglected.

Iterative coupling techniques are also applied to improve the classical IMPES scheme. For instance, in \cite{Kou2010b} an iteration   between the pressure and saturation equation is introduced. This iterative scheme is based on their previous work \cite{Kou2010}. Radu et al., \cite{Radu2015} have proposed a fixed-point iterative scheme for two-phase flow model (in global pressure formulation). 
Recently, Kvashchuk et al., \cite{Anna} have proposed an iterative linearisation scheme for two-phase flow (in average pressure formulation) following IMPES. 
The scheme approximates the capillary pressure function by applying a chain rule  and evaluating the  non-linear terms at the previous iteration. This approximates the transport equation semi-implicitly. However, the pressure equation was evaluated at the previous iteration saturation profile. This implies that the scheme lacks a coupling term at the current time step. As a consequence, the scheme might be challenged by dynamic capillary pressure forces that change the saturation distributions in a very short time.

In this paper, we propose and analyse an iterative linearization scheme for the designed non-standard model above based on iterative IMPES approach, typically we followed the work of Kvashchuk et al., \cite{Anna}. We discretize the dynamic capillary pressure functions semi-implicitly in time, where the gradient of the dynamic capillary pressure function (in the pressure and saturation equations) is reformulated by applying a chain rule (see Equation (\ref{twophase2}) in section \ref{semi-impl}). We then introduce an iteration step and evaluated the non-linear terms at the previous iteration.  We further introduce an $L$-scheme type \cite{Radu2015, List2016} stabilization term in the pressure and transport equations.  
We prove the convergence and robustness of the proposed scheme under natural assumptions.  
The convergence proof shows that the linearization technique and the  introduced stabilization terms  allowed the scheme to  take large time step size. In contrast to the classical Newton method, the proposed scheme can be seen as an inexact Newton method which has an advantage of not computing the Jacobian of the system.

This paper is organized as follows. Section \ref{NonlocalTPF} describes the mathematical model of non-standard immiscible  incompressible two-phase flow in porous media. In section \ref{Descretization}, we introduce a linearization scheme for the resulting model, and proved the convergence of  the proposed scheme. We further discuss the choice of a relaxation factor in this section.  
Numerical simulations in 2D and 3D models are presented in section \ref{NumRes}.  This section shows the performance of the proposed scheme and compares  it with iterative IMPES. The paper ends by a conclusive remark in section \ref{Con}.

\section{Non-local two-phase flow model}\label{NonlocalTPF}
Let ${\rm \Omega}$ be a bounded permeable domain in $\mathbb{R}^d$, $d = 1, 2 ~{\rm or }~3$, having a Lipschitz continuous boundary ${\rm \partial\Omega}$ and let $t \in [0,T]$ be the life time of the processes. The two-phase  flow in such domain is governed by Darcy's law and mass balance  equations for each phase \cite{Nordboton2012}. For each phase $\alpha\in \{w,o\}$, where $w$, $o$ stand for wetting and   non-wetting fluids respectively, the Darcy flux $\textbf{u}_\alpha : {\rm \Omega\times [0,T]}\rightarrow \mathbb{ R}^d$ is given by  
\begin{eqnarray}\label{Darcy}
\mathbf{u}_\alpha = - \lambda_{\alpha}\big( \nabla P_\alpha - \rho_\alpha g\nabla z\big),
\end{eqnarray}
where $\lambda_\alpha : {\rm \Omega\times [0,T]}\rightarrow \mathbb{ R}$ is phase $\alpha$ mobility, $\rho_\alpha : {\rm \Omega\times [0,T]}\rightarrow \mathbb{ R}$ is phase density that controls the buoyancy force, and $g$ is the gravitational vector. The phase mobility is defined as $\lambda_\alpha = \frac{\mathbb{K} k_{r\alpha}}{\mu_\alpha}$, where $\mathbb{K} : \mathrm{\Omega} \rightarrow \mathbb{R}^{d\times d}$ is the absolute permeability of the rock, $k_{r\alpha}$ is phase $\alpha$ relative permeability, and $\mu_\alpha$ is phase $\alpha$ viscosity.  

For each phase $\alpha\in \{w,o\}$, the balance of mass for the incompressible immiscible  fluids yield the transport equations,
\begin{eqnarray}\label{continuty}
\phi\rho_\alpha \partial_t S_\alpha +  \rho_\alpha \nabla \cdot \mathbf{u}_\alpha = f_{\alpha}, ~{\rm in}~ \mathrm{\Omega},
\end{eqnarray}
where $\phi$ is the porosity of the medium $\mathrm{\Omega}$, and $f_\alpha$ is source or sink term in each phase. From model \eqref{Darcy} and \eqref{continuty}, we obtained two equations with four unknown  variables. To close the system the following constraints must also satisfied:
\begin{eqnarray}\label{cons}
0\leq S_w, S_o \leq 1, ~S_w + S_o = 1, ~\mathrm{and}~P_o-P_w = P_c(S_w),
\end{eqnarray}
where $P_c$ is the capillary pressure that relates the phase saturation to the phase pressure difference.   Equations (\ref{Darcy})-(\ref{cons}) with appropriate initial and boundary conditions are used to describe  two-phase flow dynamics in porous medium.

\subsection{Model reformulation}
Since we are dealing with incompressible fluids and matrix, we can sum up the mass balance  models in Equation (\ref{continuty}) to get the pressure equation, 
\begin{align}\label{pressure}
-\nabla\cdot \Big(\lambda_{tot} \nabla P_o   - \lambda_w\nabla P_c- (\lambda_w\rho_w + \lambda_n\rho_w)g\nabla z \Big) = \textbf{f}_p + f_s&& ~{\rm in}~ \mathrm{\Omega},
\end{align}
where $\lambda_{tot} = \lambda_{w}+\lambda_{o}$ is the total mobility. 
In Equation (\ref{pressure}), we have one equation and two unknowns namely, $P_o$ and $S_w$. As a consequence, the transport equation for the wetting or non-wetting phase saturation should come to play in order to close the system.  Therefore, we get a system of two equations with two unknowns,

\begin{subequations}
 \begin{align}\label{pressure_p}
-\nabla\cdot \Big(\lambda_{tot} \nabla P_o   - \lambda_w\nabla P_c- (\lambda_w\rho_w + \lambda_n\rho_n)g\nabla z \Big) = f_t  & &{\rm in}~ \mathrm{\Omega},\\
\phi\partial_t S_w - \nabla \cdot \lambda_w\Big( \nabla P_o - \nabla P_{c} - \rho_wg\nabla z\Big) = f_{s}& &{\rm in}~ \mathrm{\Omega},\label{pressure_s}
\end{align}
\end{subequations}
where, $f_t = f_w+f_n$ is the total source. In order  to solve the two Equations (\ref{pressure_p}) and (\ref{pressure_s}), one needs to impose appropriate initial and boundary conditions, such as Neumann and Dirichlet conditions. Thus, we assume that the boundary of the system is divided into disjoint sets such that $\partial\mathrm\Omega = \Gamma_D \cup \Gamma_N$. We denote by $\nu$ the outward unit vector normal to $\partial\mathrm{\Omega}$, and set
\begin{flalign*}
& P_o(\cdot, 0) = P_o^0(\cdot), & & S_w(\cdot, 0) = S_w^0(\cdot), & & {\rm in}~\mathrm\Omega,\nonumber\\
& P_o  = P_{o,D},& & S_w  = S_{w,D},& &{\rm on} ~\mathrm\Gamma_D \times (0,T],\nonumber\\
& \textbf u_\alpha  = J_\alpha, && &&{\rm on} ~\mathrm\Gamma_N \times (0,T],\nonumber
\end{flalign*}
where $J_\alpha\in \mathbb R^d$ is phase inflow rate. In order to make the model uniquely determined, it is required that $\mathrm\Gamma_d\neq\emptyset$.
\subsection{Relative permeability and dynamic capillary pressure functions}\label{relacap}

Commonly, the Brooks-Corey \cite{RefBrooksCorry} and van Genuchten \cite{RefvanGenuchten} models are used to represent the capillary pressure and relative permeabilities for equilibrium system. For non-equilibrium systems, explicit time-dependency of \pcs curves have been developed (for example see \cite{RefHassanizadeh, RefDahle}) to capture changes in capillary pressure induced by dynamic flow conditions. These models are developed under static wettability condition. 

In this paper, we consider an extended capillary pressure model that captures the dynamic change of rock wettability at pore-scale. 
Kassa et al., \cite{Kassa2018}  have introduced the dynamic term   as an interpolation between the end wetting state curves. This can be described mathematically as follows, 
\begin{eqnarray}
 P_c=(1-\omega(\cdot))P_c^{ ww} + \omega(\cdot) P_c^{ow},
    \label{eq:dynPc_interp} ~~~~~~~~~~~~~~
\end{eqnarray}
where, $P_c^{ ww}$ and $P_c^{ ow}$ are end wetting (respectively the water-wet and oil-wet) capillary pressure functions. Here, the water-wet and oil-wet capillary pressure functions are represented  respectively with large  and small (possibly negative) entry pressures. The dynamic coefficient  $\omega(\cdot)$ is designed to capture the dynamics of time-dependent WA at the pore-scale. In \cite{Kassa2018}, time-dependent  pore-scale  WA processes are upscaled to Darcy scale to represent the coefficient term in the extended capillary pressure model (\ref{eq:dynPc_interp}). The upscaled dynamic coefficient $\omega(\cdot)$ is governed by  Langmuir adsorption type models,
\begin{equation}\label{omega}
\omega(S_w, t) = \left\{\begin{array}{l}
\frac{\beta_1\chi}{\beta_1\chi + 1}, ~~~~{\rm for ~uniform ~alteration}\\[0.1in]
\frac{\beta_2S_w\chi}{\beta_2S_w\chi + 1}, ~{\rm for ~non-uniform ~alteration}\
\end{array}\right. \chi :=\frac{1}{T} \int_0^t (1-S_{w})d\tau,
\end{equation}
where $\beta_1$ and $\beta_2$ are fitting parameters that are dependent on the pore-scale WA model parameter (see the details in \cite{Kassa2018}), $T$ is the characteristic time and we recommend to choose $T$ such that $\chi\in [0,1]$. 

The non-linear variable $\chi$  captures the effect of pore-scale time-dependent WA dynamics   for the exposed porous representative elementary volume (REV), and it is an increasing function of time. Thus, the models (\ref{eq:dynPc_interp})-(\ref{omega}) describe time-dependent WA induced dynamic capillary pressure model. The change of the capillary pressure in time continues  for constant water saturation. However, $\chi$ keeps constant for pores that are fully occupied with water, i.e., $S_w=1$. In this case, the capillary pressure is only dependent on the current water saturation path. This may lead to discontinuity of the capillary pressure function at the interface of grid blocks. Thus, the continuity of capillary pressure results in  a saturation discontinuity.  

For end  wetting (water-wet and oil-wet) conditions, we considered two consistent set of capillary pressure functions. Qualitatively, these curves represent either water-wet ($ww$) or oil-wet ($ow$) conditions. We adopted the van Genuchten constitutive model for these  conditions and can be read as, 
\begin{eqnarray}\label{cap-p}
 P^{ww}_c = P_e^w\Big(S^{-\frac{1}{m_w}}_w-1\Big)^\frac{1}{n_w},~{\rm and}~~P^{ow}_c = P_e^o\Big(S^{-\frac{1}{m_o}}_w-1\Big)^\frac{1}{n_o},
\end{eqnarray}
where $P_e^\alpha$ is phase $\alpha$ wetting condition entry pressure, $m_\alpha$ is pore volume distribution of the porous domain for the $\alpha$'s wetting condition and can be related to $n_\alpha$ as $m_\alpha = 1/n_\alpha$.
In this study, only the standard van Genuchten relative permeability functions are considered to describe the relative movement of fluids,
\begin{align}\label{caplam}
k_{r\alpha}^{ww} = \left\{\begin{array}{l r}\sqrt{1-S_w}\Big(1-S_w^\frac{1}{m_w}\Big)^{2m_w}, & \alpha=o, \\\
\sqrt{S_w}\Big(1-\big(1-S_w^\frac{1}{m_w}\big)^{m_w}\Big)^2, &\alpha=w.\\
\end{array}\right.  
\end{align}

Coupling the relations $k_{r\alpha}^{ww}$-$S_w$ and \pcs-$\chi$  into  the flow model (\ref{pressure_p})-(\ref{pressure_s}) will give non-standard dynamic two-phase flow model in porous media. The goal of this study is to propose a stable and flexible scheme that handles such dynamics efficiently for simulations that consider long-term time evolution.    
\section{Discretisation, linearisation and iterative coupling technique}\label{Descretization}
%

%
Let the total simulation time interval $[0,T]$ be divided into $N$ time steps in such a way that $0=t^0<t^1< \cdots < t^N = T$, and define the time step $\delta t = T/N$, as well as $t^n = n\delta t,~ n \in \{1,2,\ldots, N\}$.

Backward Euler method is applied to discretize the resulting non-local two-phase flow model in time and the semi-discretized model can be read as,
\begin{subequations}
\begin{eqnarray}\label{pressure_d}
-\nabla\cdot \Big(\lambda_{tot}(S_w^{n+1}) \nabla  P_o^{n+1} -   \lambda_w(S_w^{n+1})\nabla P_c(\chi^{n+1},S_w^{n+1})  \Big)  =   f_{t}^{n },\\
\phi\frac{S_w^{n+1}-S_w^n}{\delta t}   -  \nabla \cdot \Big(\lambda_w(S_w^{n+1})\Big( \nabla P_o^{n+1} -  \nabla P_c(\chi^{n+1},s_w^{n+1})\Big)\Big) = f_{s} ^{n }.\label{continuty_D}
\end{eqnarray}
\end{subequations}
The superscripts ($n$+1) and $n$ represent the current and previous time steps respectively.  Above, we omitted the gravity term and the analysis will continue in this form for the sake of clarity and brevity of the presentation.  

The above system is fully coupled and challenging to  solve directly because of its non-linearity. Due to the non-linearity, iterative linearization and sequential coupling  methods such as iterative IMPES are needed to solve such systems.
\subsection{Iterative IMPES}
The iterative IMPES linearises the given two phase flow problem by evaluating the non-linear terms from the previous iteration step \cite{Kou2010b}. Thus, the non-linear model (\ref{pressure_d})-(\ref{continuty_D}) can be reduced to,
\begin{subequations}
\begin{eqnarray}\label{pressureD}
-\nabla\cdot \Big(\lambda_{tot}(S_w^{n+1,i}) \nabla  P_o^{n+1,i+1} -   \lambda_w(S_w^{n+1})\nabla P_c(\chi^{n+1,i},S_w^{n+1,i})  \Big)  =   f_{t}^{n }, ~~~~~\\ 
\phi\frac{S_w^{n+1,i+1}-S_w^n}{\delta t}   -  \nabla \cdot \Big(\lambda_w(S_w^{n+1,i}) \Big(\nabla P_o^{n+1,i+1} -  \nabla P_c(\chi^{n+1,i},S_w^{n+1,i})\Big)\Big) = f_{s} ^{n }.~~~~~\label{continutyD}
\end{eqnarray}
\end{subequations}
The iterative IMPES solver starts with $S_w^{n+1,i} = S_w^{n}$ and thus, the system above is linear and decoupled. Usually the pressure equation (\ref{pressureD}) is solved for $P_o^{n+1,i+1}$ first. The computed pressure and the previous iteration saturation profile are used to update the current iteration saturation profile explicitly from Equation (\ref{continutyD}). The iteration will continue until the convergence criterion has been satisfied.


%

\subsection{Semi-implicit time discretization}\label{semi-impl}
The iterative IMPES formulation above splits the pressure and saturation equations in each iteration step. Hence, the approach has missed the inherent coupled nature of the original problem (\ref{pressure_d})-(\ref{continuty_D}). This may lead to instability on the convergence of the method in particular for long-term reservoir processes.  

In this paper, we propose a scheme that couples the pressure, and saturation equations at the ($n$+1)-th time step in addition to the current iteration step. The scheme treats the dynamic capillary pressure function (in the pressure and saturation equations) semi-implicitly in time. We then introduce a monotone  fixed-point iteration \cite{Radu2015, Anna, List2016, Sorin2004, Kou2010b}. The development of the scheme is discussed below.

The scheme starts with approximating the capillary pressure function at the current time step (in the pressure  (\ref{pressure_d}) and saturation (\ref{continuty_D}) equations) by applying chain rule and semi-backward Euler discretization in time. The resulting approximation is read as,
\begin{equation}\label{twophase2}
\nabla P_c^{n+1} \approx \frac{\partial P_c^n}{\partial S_w} \nabla S_w^{n+1} + \frac{\partial P_c^n}{\partial \chi}\nabla \chi^{n+1}.
\end{equation}
The obtained approximate capillary pressure is substituted back to the two-phase flow model to give the following linear system (we call this linearization technique \textit{pseudo-monolithic} scheme),
\begin{subequations}
\begin{eqnarray}\label{pressureNew}
-\nabla\cdot \Big(\lambda_{tot}^{n} \nabla P_o^{n+1} - \lambda_w^{n}\Big(\frac{\partial P_c^{n}}{\partial S_w} \nabla S_w^{n+1} +  \frac{\partial P_c^{n}}{\partial \chi}\nabla \chi^{n+1} \Big)\Big)  =   f_{t}^{n},~~~~~\\ 
\phi\frac{S_w^{n+1}-S_w^n}{\delta t}   -  \nabla \cdot \Big(\lambda_w^{n} \Big(\nabla P_o^{n+1}  -  \Big(\frac{\partial P_c^{n}}{\partial S_w} \nabla S_w^{n+1} + \frac{\partial P_c^{n}}{\partial \chi}\nabla \chi^{n+1} \Big)\Big)\Big) = f_{s}^{n}.~~~~~\label{continutyNew} 
\end{eqnarray}
\end{subequations}
The above approach (\ref{pressureNew})-(\ref{continutyNew}) couples the pressure and saturation equations at the current time step weakly. But, importantly, the saturation and pressure state variables communicate  each other at the same degree of decision making level. Recall that the variable $\chi$ is also  a function of saturation, and thus, the number of equations and unknowns are compatible. 

Then stability and accuracy of the pseudo-monolithic scheme (\ref{pressureNew})-(\ref{continutyNew}) is improved further by introducing outer iteration steps (i.e., ($i+1$) and $i$), and evaluating the non-linear terms at the current time step ($n$+1 instead of $n$) but at the previous iteration $i$. We controlled the convergence of the proposed fixed point iteration by adding an $L$-scheme type \cite{Radu2015, List2016, Kou2010b} stabilization term.  We named this linearization technique as \textit{iterative pseudo-monolithic} scheme, and read as, 
\begin{subequations}
\begin{eqnarray}\label{pressureNew2}
-\nabla\cdot \Big(\lambda_{tot}^{n+1,i} \nabla  P_o^{n+1,i+1} - ~~~~~~~~~~~~~~~~~~~~~~~~~~~~~~~~\nonumber\\ \lambda_w^{n+1,i}\Big(\frac{\partial P_c^{n+1,i}}{\partial S_w} \nabla  \widetilde{S}_w^{n+1,i+1} +  \frac{\partial P_c^{n+1,i}}{\partial \chi}\nabla \widetilde{\chi}^{n+1,i+1} \Big)\Big)  =   f_{t}^{n },~~~\\ 
\phi\frac{\widetilde{S}_w^{n+1,i+1}-S_w^n}{\delta t}   -  \nabla \cdot \Big(\lambda_w^{n+1,i} \Big(\nabla P_o^{n+1,i}  - ~~~~~~~~~~~~~~~~~~~~~~~~~~~~~~~~\nonumber\\  \Big(\frac{\partial P_c^{n+1,i}}{\partial S_w} \nabla  \widetilde{S}_w^{n+1,i+1} + \frac{\partial P_c^{n+1,i}}{\partial \chi}\nabla \widetilde{\chi}^{n+1, i+1} \Big)\Big)\Big) = f_{s}^{n },~~~ \label{continutyNew2} \\
~S_w^{n+1,i+1}  = (1-L^{i+1})S_w^{n+1,i} + L^{i+1} \widetilde{S}_w^{n+1,i+1},~~~~~~~~~~~~~~~~~~~\label{relaxationfa_1}\\
~\chi^{n+1,i+1} = (1-L^{i+1})\chi^{n+1,i} + L^{i+1} \widetilde{\chi}^{n+1,i+1},~~~~~~~~~~~~~~~~~~~\label{relaxationfa}
 \end{eqnarray}
\end{subequations}
where $L^{i+1} \in (0,1]$ is a stabilization constant that has an important role on the convergence of the proposed scheme. The choice of $L^{i+1}$ in each iteration will be discussed later in this paper. Equations (\ref{relaxationfa_1}) and (\ref{relaxationfa}) can be substituted into Equations (\ref{pressureNew2}) and (\ref{continutyNew2}) directly during the solution processes. Here, we note that $S_w^{n+1,i+1}$ and $\chi^{n+1,i+1}$ are used as a previous iteration values for the next iteration and we set $S_w^{n+1,0} = S_w^{n+1}$ for the first iteration step. We also note that $\chi$ is a function of saturation. The terms in Equations (\ref{pressureNew2})-(\ref{relaxationfa}) are linear and coupled in each iteration. 

\begin{remark}
The pseudo-monolithic and the iterative pseudo-monolithic schemes reduced to IMPES and iterative $L$-scheme respectively if the capillary pressure is zero or neglected. 
\end{remark}

Below, we demonstrate the convergence of the iterative pseudo-monolithic scheme, and in section \ref{NumRes}, we compare its performance  against  the pseudo-monolithic scheme (\ref{pressureNew})-(\ref{continutyNew}) and the iterative IMPES. 
\subsubsection{Convergence analysis of the iterative pseudo-monolithic scheme}\label{con}
%
%

We denote by $L_2(\mathrm{\Omega})$ the space of real valued square integrable functions, and by $H^1(\mathrm{\Omega})$ its subspace containing functions having also the first order derivatives in $L_2(\mathrm{\Omega})$. Let $H_0^1(\mathrm{\Omega})$ be the space of functions in $H^1(\mathrm{\Omega})$ which vanish on the boundary. Further, we denote by $\langle\cdot ,\cdot\rangle$ the inner product on $L_2(\mathrm{\Omega})$, and by $\| \cdot \|$ the norm of $L_2(\mathrm{\Omega})$. $L_f$ stays
for the Lipschitz constant of a Lipschitz continuous function
$f(\cdot)$.

Let $T_h$ is a regular decomposition of $\mathrm{\Omega}$, which decomposes $\mathrm{\Omega}$ into closed $d$-simplices; $h$ stands for the mesh diameter. Here we assume $\mathrm{\Omega} = \bigcup_{ \mathcal{T} \in  \mathcal{T}_h} \mathcal{T}$. The Galerkin finite element space is given by
\begin{equation} 
\textbf{V}_h :=\Big\{\textbf{v}_h \in H_0^1 (\Omega) \mid \textbf{v}_h|_ \mathcal{T} \in P_1( \mathcal{T}),  \mathcal{T} \in  \mathcal{T}_ h \Big\},
\end{equation}
where $P_1 ( \mathcal{T})$ denotes the space of linear polynomials on any simplex $T$.

We use the definition of spaces and notations above to write the the variational form of Equations (\ref{pressure_d})-(\ref{continuty_D}) which finds $P_o^{n+1},S_w^{n+1}\in\textbf{ V}_h$ for a given $S_w^n$  such that the following holds, 
\begin{subequations}
\begin{eqnarray}\label{varequap}
\Big\langle \lambda_{tot}^{n+1} \nabla  P_o^{n+1}  -~~~~~~~~~~~~~~~~~~~~~~~~~~~~~~~~~~~~~~~~~~~~~~~~~~~~~~~~~~~~\nonumber\\ \lambda_w^{n+1}\Big(\frac{\partial P_c^{n+1}}{\partial S_w} \nabla S_w^{n+1} +  \frac{\partial P_c^{n+1}}{\partial \chi}\nabla \chi^{n+1} \Big), \nabla v_h\Big\rangle = \Big\langle f_t^n , v_h\Big\rangle~~ \\
\Big\langle \phi S_w^{n+1}-\phi S_w^{n+1}, \textbf{v}_h\Big\rangle +
\delta t\Big\langle \lambda_w^{n+1} \nabla P_o^{n+1} -  ~~~~~~~~~~~~~~~~~~~~~~~~~~\nonumber\\\lambda_w^{n+1}\Big(\frac{\partial P_c^{n+1}}{\partial S_w} \nabla S_w^{n+1} +  \frac{\partial P_c^{n+1}}{\partial \chi}\nabla \chi^{n+1} \Big), \nabla \textbf{v}_h\Big\rangle = \Big\langle  f_{s}^{n}, \textbf{v}_h\Big\rangle,~~ \label{varequap-2}
\end{eqnarray}
\end{subequations}
for all $v_h \in \textbf{V}_h$. 
Similarly, we can also write the variational form of the iterative pseudo-monolithic method (\ref{pressureNew2})-(\ref{continutyNew2}) that  find  $P_o^{n+1,i+1}, S_w^{n+1,i+1} \in \textbf{V}_h$ for given $S_w^n, ~S_w^{n+1,i}$   such that
\begin{subequations}
\begin{eqnarray}\label{varequa2}
\Big\langle \lambda_{tot}^{n+1,i} \nabla  P_o^{n+1,i+1} - ~~~~~~~~~~~~~~~~~~~~~~~~~~~~~~~~~~~~~~~~~~~~~~~~~~~~~~~~~~~~~~~~~~~~~~~~~~~~~~\nonumber\\\lambda_w^{n+1,i}\Big(\frac{\partial P_c^{n+1,i}}{\partial S_w} \nabla \widetilde{S}_w^{n+1,i+1} +  \frac{\partial P_c^{n+1,i}}{\partial \chi}\nabla \widetilde\chi^{n+1,i+1} \Big), \nabla v_h\Big\rangle = \Big\langle f_t^{n } , v_h\Big\rangle ~~~~~~~ \\
 \Big\langle \phi \widetilde{S}_w^{n+1,i+1}- \phi S_w^n,  v_h\Big\rangle + \delta t\Big\langle \lambda_w^{n+1,i} \Big(\nabla P_o^{n+1,i+1} -  ~~~~~~~~~~~~~~~~~~~~~~~~ ~~~~~~~~~~~~~~~~~\nonumber\\\Big(\frac{\partial P_c^{n+1,i}}{\partial S_w} \nabla \widetilde{S}_w^{n+1,i+1} +  \frac{\partial P_c^{n+1,i}}{\partial \chi}\nabla \widetilde\chi^{n+1,i+1} \Big)\Big), \nabla v_h\Big\rangle = \Big\langle  f_{s}^{n }, v_h\Big\rangle,~~~~~~ \label{varequa2-s}
\end{eqnarray}
\end{subequations}
holds for all $v_h\in \textbf{V}_h$. The aim is to show that the linearized model (\ref{varequa2})-(\ref{varequa2-s}) converges to the non-linear problem (\ref{varequap})-(\ref{varequap-2}) within few outer iteration steps in each time step. 
%


The convergence analysis of the scheme is proved theoretically by assuming that the continuous model has a solution.  Further, the following assumptions on the coefficient functions and the discrete solutions are defining the framework in which we can prove the convergence of the proposed scheme. 
\begin{itemize}
\item[\textit{A1}:] The mobilities satisfy the Lipschitz continuity condition in the wetting phase saturation, i.e., there exist constants $L_{\lambda_\alpha}$ such that 
\begin{equation}
\Vert\lambda_{\alpha}(S_w) - \lambda_{\alpha}(\overline{S}_w)\Vert \leq L_{\lambda_\alpha} \Vert S_w -\overline{S}_w\Vert, ~~ \forall S_w, \overline{S}_w \in [0,1]. 
\end{equation}
This implies that any linear combination of $\lambda_\alpha$ is also Lipschitz continuous. 
\item[\textit{A2}:]  The dynamic capillary pressure function $P_{c}$, and its partial  derivatives $\frac{\partial P_{c}}{\partial S_w}$ and $\frac{\partial P_{c}}{\partial \chi}$ are  Lipschitz continuous with respect to $S_w$ and $\chi$. This implies, for any $\chi,\overline{\chi}, S_w, \overline{S}_w \in [0,1]$, we can find constants $L_{P_c}^\chi, L_{P'_c}^\chi,   L_{P_c}^s~{\rm and} ~L_{P_c'}^s$  such that 
\begin{eqnarray}
\Vert P_c(\chi,S_w) - P_c(\chi,\overline{S}_w)\Vert \leq L^s_{P_c}\Vert S_w - \overline{S}_w\Vert, ~{\rm and}~~~~~~~~\nonumber \\ 
\Big\Vert \frac{\partial P_c(\chi,S_w) }{\partial S_w}- \frac{\partial P_c(\chi,\overline{S}_w)}{\partial S_w}\Big\Vert \leq L^s_{\Phi_c'}\Vert S_w - \overline{S}_w\Vert,
\end{eqnarray}
\begin{eqnarray}
\Vert P_c(\chi,S_w) - P_c(\overline{\chi},S_w)\Vert \leq L_{\Phi_c}^\chi\Vert \chi- \overline{\chi}_w\Vert, ~{\rm and}~~~~~~~~\nonumber \\ 
\Big\Vert \frac{\partial P_c(\chi,S_w)}{\partial \chi} - \frac{\partial P_c(\overline{\chi},S_w)}{\partial \chi}\Big\Vert \leq L^\chi_ {P_c'}\Vert \chi- \overline{\chi}\Vert.
\end{eqnarray}
Further, we assume that the dynamic capillary pressure $P_c(\chi,S_w)$ is decreasing function, i.e., $\frac{\partial P_c(\chi,S_w)}{\partial S_w} < 0$, and $\frac{\partial P_c(\chi,S_w)}{\partial \chi} <0$ 

\item[\textit{A3}:] We assumed that the initial wetting phase saturation   satisfies $\| \nabla S_w^{n}\|_\infty \leq M_{s}$  with $\|\cdot\|$ denoting the $L^\infty (\mathrm\Omega)$-norm. This implies also $\| \nabla S_w^{n+1}\|_\infty \leq M_{s}$ and $\| \nabla P_n^{n+1}\|_\infty \leq M_{p}$. 

\item[\textit{A4}:] The total derivative of $P_c$ with respect to $S_w$ is bounded above by zero. 

\item[\textit{A5}:] Assume that for any time step ($n$+1) with $n\geq 0$, there exist a solution for saturation $S_w^{n+1}$ and pressure $P_o^{n+1}$ such that the Equations (\ref{varequap})-(\ref{varequap-2}) are satisfied.
\end{itemize}
From now on, we denote by 
\begin{equation}
\textbf{e}_p^{i+1} = P_o^{n+1,i+1} - P_o^{n+1}, ~ \textbf{e}_s^{i+1} = S_w^{n+1} -  S_w^{n+1,i+1}, ~\widetilde{\textbf{e}}_s^{i+1} = S_w^{n+1} - \widetilde{S}_w^{n+1,i+1}, 
\end{equation}
the error at iteration $i+1$. A scheme is convergent if $\| \textbf{e}_p^{i+1}\|\rightarrow 0,$ $\| \textbf{e}_s^{i+1}\|\rightarrow 0$
when $i\rightarrow\infty$.
\begin{theorem}
Assume that the conditions (A1)-(A5) are satisfied.  If we choose $S_w^n$  as the initial approximation, $S_w^{n+1,0}$, of the exact solution $S_w^{n+1}$, there exist a time step size $\delta t^n$ with mild restriction such that the iteration $S_w^{n+1,i+1}$ and $P_o^{n+1,i+1}$ generated by the scheme (\ref{varequa2})-(\ref{varequa2-s}) converges to $S_w^{n+1}$ and $P_o^{n+1}$ respectively in $L_2$ norm.  
\end{theorem}
\begin{proof}: As in \cite{Anna,List2016,Radu2015}, we start the analysis by subtracting the linearized  pressure equation (\ref{varequa2}) from non-linear Equation  (\ref{varequap}) to obtain:
\begin{eqnarray}\label{varequaP3}
\Big\langle  \lambda_{tot}^{n+1} \nabla P_o^{n+1} - \lambda_{tot}^{n+1,i} \nabla P_o^{n+1,i+1} , \nabla v_h\Big\rangle ~~~~~~~~~~~~~~~~~~~~~~~~~~~~~~~~~~~~~~~~\nonumber\\ -
 \Big\langle \lambda_w^{n+1}    \frac{\partial P_c^{n+1}}{\partial S_w}\nabla S_w^{n+1} -  \lambda_w^{n+1,i}  \frac{\partial P_c^{n+1,i}}{\partial S_w}\nabla \widetilde S_w^{n+1,i+1}, \nabla v_h\Big\rangle~~~~~~~~~~~~~~~~ \nonumber\\ 
-\Big\langle \lambda_w^{n+1}    \frac{\partial P_c^{n+1}}{\partial \chi}\nabla \chi^{n+1} -  \lambda_w^{n+1,i}  \frac{\partial P_c^{n+1,i}}{\partial \chi}\nabla \widetilde \chi^{n+1,i+1}, \nabla v_h\Big\rangle= 0,~~~~~
\end{eqnarray}
for any $v_h\in V_h$.  Applying the Cauchy-Schwartz inequality 
\begin{equation}
\vert\langle u, v\rangle\vert^2 \leq \Vert u\Vert^2\Vert v \Vert^2, 
\end{equation}
followed by the assumptions (A1)-(A4), and testing with $v_h =\textbf{ e}_p^{i+1}$, we get the following estimate,
\begin{eqnarray}\label{varequaP5}
M_{\lambda_{tot}}^0\Big\langle \nabla \textbf{e}_{p}^{i+1},\nabla \textbf{e}_p^{i+1}\Big\rangle \leq\gamma_p\| \textbf{e}_{s}^i\|\|\nabla \textbf{e}_p^{i+1}\|  +~~~~~~~~~~~~~~~~~~~~~~~~~~~\nonumber\\ \Big\langle\lambda_w^{n+1,i}  \Big(\frac{\partial P^{n+1,i}_c}{\partial S_w}-\frac{t}{T}\frac{\partial P^{n+1,i}_c}{\partial \chi}\Big)\nabla \widetilde{\textbf{e}}_s^{i+1}, \nabla \textbf{e}_p^{i+1}\Big\rangle,
\end{eqnarray}
where   $\gamma_p = L_{\lambda_{tot}}M_p + L_{\lambda_w} (M^s_{P_c} + M^\chi_{P_c}) + \big(L_{P'_c}^s+L_{P'_c}^\chi\big) M_{\lambda_w}$. Applying (A4) once more, we obtain an estimate as follows,  
\begin{eqnarray}\label{varequaP6}
\|\nabla \textbf{e}_{p}^{i+1}\| \leq  \frac{\gamma_p}{M_{\lambda_{tot}}^0} \| \textbf{e}_{s}^i\| 
\end{eqnarray}

Similarly, we subtract  Equation   (\ref{varequa2}) from Equation (\ref{varequap-2}) to get,
\begin{eqnarray}\label{eqvasatu}
\frac{\phi}{\delta t}\Big\langle \widetilde{\textbf{e}}_s^{i+1},v_h\Big\rangle 
-  \Big\langle \lambda_w^{n+1}    \frac{\partial P_c^{n+1}}{\partial S_w}\nabla S_w^{n+1} - \lambda_w^{n+1,i}  \frac{\partial P_c^{n+1,i}}{\partial S_w}\nabla S_w^{n+1}~~~~~~~~~~~~~~~ \nonumber\\
+ \lambda_w^{n+1,i}  \frac{\partial P_c^{n+1,i}}{\partial S_w}\nabla S_w^{n+1}-
  \lambda_w^{n+1,i}  \frac{\partial P_c^{n+1,i}}{\partial S_w}\nabla 	\widetilde{S}_w^{n+1,i+1}, \nabla v_h\Big\rangle \nonumber\\  
-   \Big\langle \lambda_w^{n+1} \frac{\partial P_c^{n+1}}{\partial \chi}\nabla \chi^{n+1} - \lambda_w^{n+1,i}  \frac{\partial P_c^{n+1,i}}{\partial \chi}\nabla \chi^{n+1} + ~~~~~~~~~~~\nonumber\\ 
\lambda_w^{n+1,i}  \frac{\partial P_c^{n+1,i}}{\partial \chi}\nabla \chi^{n+1} - 
 \lambda_w^{n+1,i}  \frac{\partial P_c^{n+1,i}}{\partial \chi}\nabla \widetilde{\chi}^{n+1,i+1}, \nabla v_h\Big\rangle \nonumber \\  
-  \Big\langle \lambda_{w}^{n+1,i}(\nabla P_o^{n+1,i+1} - \nabla P_o^{n+1}),\nabla v_h\Big\rangle + ~~~~~~~~~~~~~~~~~\nonumber\\  \Big\langle (\lambda_{w}^{n+1,i} - \lambda_{w}^{n+1}) \nabla P_o^{n+1}, \nabla v_h\Big\rangle = 0.~~~~~~~~
\end{eqnarray}
Now by taking the advantage of assumptions (A1)-(A3) and  applying the Cauchy-Schwartz inequality with the definition of $\chi$, Equation (\ref{eqvasatu}) can be estimated as,
\begin{eqnarray}\label{eqvasatu2s}
\frac{\phi}{\delta t}\| \widetilde{\textbf{e}}_s^{i+1}\|  
- \Big\langle\lambda_w^{n+1,i} \Big( \frac{\partial P_c^{n+1,i}}{\partial s_w} - \frac{t}{T}\frac{\partial P_c^{n+1,i}}{\partial \chi}\Big)\nabla \widetilde{\textbf{e}}_s^{i+1}, \nabla \widetilde{\textbf{e}}_s^{i+1}\Big\rangle   \nonumber\\ 
   \leq  M_{\lambda_w}\|\nabla   \textbf{e}_{p}^{i+1}\|\|\nabla \widetilde{\textbf{e}}_s^{i+1}\| + L_{\lambda_w}M_p\|\textbf{e}_s^{i}\|\|\nabla \widetilde{\textbf{e}}_s^{i+1}\|,
\end{eqnarray}
where we choose $v_h = \widetilde{\textbf{e}}_s^{i+1}$ as a test function. 
At this point, we apply assumption (A1) and (A4). From assumption (A4), we have that $ \frac{\partial P_c^{n+1,i}}{\partial S_w} - \frac{t}{T}\frac{\partial P_c^{n+1,i}}{\partial \chi} < 0$. This implies that there exists a real number $M_{P'_c}> 0$ such that, \[\max_{S_w,t}\Big\{ \frac{\partial P_c^{n+1,i}}{\partial S_w} - \frac{t}{T}\frac{\partial P_c^{n+1,i}}{\partial \chi}\Big\} = -M_{P'_c}.\] Considering all these and after some algebraic manipulation, the inequality \eqref{eqvasatu2s} can be rewritten as
\begin{eqnarray}\label{eqvasatu21}
\frac{\phi}{\delta t}\|  \widetilde{\textbf{e}}_s^{i+1}\|^2
  +  M_{P'_c}   \| \nabla  \widetilde{\textbf{e}}_s^{i+1}\|^2 \leq  M_{\lambda_w}\| \nabla  \widetilde{\textbf{e}}_{p}^{i+1}\|\|\nabla \widetilde{\textbf{e}}_s^{i+1}\|  + \gamma_s \|\textbf{e}_{s}^i\| \|\nabla \widetilde{\textbf{e}}_s^{i+1}\|.
\end{eqnarray} 
where $\gamma_s = L_{\lambda_w}M_p$.
%
%
%
%
%
%
%
%
%
Substitute the pressure estimate (\ref{varequaP6}) into (\ref{eqvasatu21}) to give an estimate for the saturation error:
\begin{eqnarray}\label{eqvasatu5}
\frac{\phi}{\delta t}\|  \widetilde{\textbf{e}}_s^{i+1}\|^2
  +   M_{P'_c}   \| \nabla  \widetilde{\textbf{e}}_s^{i+1}\|^2 \leq  \Big(\frac{\gamma_pM_{\lambda_w}}{M_{\lambda_{tot}}^0}  + \gamma_s \Big)\| \textbf{e}_{s}^i\| \|\nabla \widetilde {\textbf{e}}_s^{i+1}\|.
\end{eqnarray}
Let us define 
\begin{equation}
C =  \frac{\gamma_pM_{\lambda_w}}{M_{\lambda_{tot}}^0}  + \gamma_s > 0,
\end{equation} 
and apply Young's inequality 
$$ab \leq \frac{a^2}{2\epsilon} + \frac{\epsilon b^2}{2},$$
for $\epsilon>0$ to the inequality (\ref{eqvasatu5}), and choosing  the parameter $\epsilon$ to be $\epsilon = \frac{C}{M_{P'_c}}$, the estimate   (\ref{eqvasatu5}) is reduced to
%
%
 %
\begin{eqnarray}\label{eqvasatu8}
\Vert  \widetilde{\textbf{e}}_s^{i+1}\Vert^2  \leq \frac{{\delta t}C^2}{M_{P'_c} \phi}  \|   \textbf{e}_{s}^i\|^2.  
\end{eqnarray}
At this stage, we can substitute the stabilization term from Equation (\ref{relaxationfa_1}) into equation (\ref{eqvasatu8}) to get the following estimate, 
\begin{eqnarray}\label{eqvasatu9}
\Big\Vert  \frac{1}{L^{i+1}}\textbf{e}_s^{i+1} +  (1-\frac{1}{L^{i+1}})\textbf{e}_s^{i}\Big\Vert^2  \leq \frac{{\delta t}C^2}{M_{P'_c} \phi}  \Big\|   \textbf{e}_{s}^i\Big\|^2.  
\end{eqnarray}
For any choice of $L^{i+1}\in (0,1]$, $1-\frac{1}{L^{i+1}}\leq 0$, and thus, by applying the reverse triangle inequality, we can obtain,
\begin{eqnarray}\label{eqvasatu10}
\Big\Vert  \textbf{e}_s^{i+1}\Big\Vert^2  \leq \Big(L^{i+1} - 1 + \frac{{\delta t}C^2}{M_{P'_c} \phi} \Big) \Big\|   \textbf{e}_{s}^i\Big\|^2.  
\end{eqnarray}
Thus, the scheme converge linearly for the designed non-local two-phase flow  model when 
\begin{equation}\label{lconver}
\delta t \leq \frac{ \big( 2-L^{i+1}\big) M_{P_c' }\phi}{C^2}   ,
\end{equation}
is satisfied.
\end{proof} 
\begin{remark}
If we choose a small $L$, convergence of the scheme is guaranteed for large time step. However, the rate of convergence   may be slow and thus, we may encounter large number of iterations. 
\end{remark}
\subsubsection{Choice of the relaxation factor}\label{relaxation}
Above we observed that the choice of the relaxation factor plays an important role on the convergence of the scheme. Here, we introduce a choice strategy for the relaxation factor based on the history of the errors at previous and current iterations.

Following \cite{Kou2010b}, we define the length of the residual of the transport equation at the current iteration by 
\begin{eqnarray}\label{resid}
\|R_s^{n+1,i+1}\| = \|\widetilde S_w^{n+1,i+1} - S_w^{n+1,i}   \|.
\end{eqnarray}
The aim is finding a relaxation factor that makes  (\ref{resid}) sufficiently small. However, this problem is highly non-linear optimization problem and thus, challenging to  come up with optimal global solution. As a consequence, we compute and bound the relaxation factor adaptively in each iteration.

To support the convergence of the iterative pseudo-monolithic scheme, the
relaxation factor $L$ should be chosen such that the residual defined by (\ref{resid}) is decreasing with each successive iterations, i.e.,
\begin{eqnarray}\label{resid2}
\|R_s^{n+1,i+1}\| \leq 
\|R_s^{n+1,i}\|
\end{eqnarray}
From (\ref{relaxationfa_1}) and (\ref{resid2}), there exists a constat $L$ such that, 
\begin{equation}\label{resid3}
  \|  S_w^{n+1,i+1} - S_w^{n+1,i} \| \leq L\|S_w^{n+1,i} - S_w^{n+1,i-1}   \|.
\end{equation}
 We denote the relaxation factor at the $i$-th iteration step by $L^i$, and in this the relaxation equation for wetting phase saturation can be rewritten as, 
\begin{equation}\label{relaxationfac2}
S_w^{n+1,i+1} = (1-L^{i+1})S_w^{n+1,i} + L^{i+1} \widetilde{S}_w^{n+1,i+1}, 
\end{equation}
where $L^{i+1}\in (0,1]$. Substituting Equation (\ref{relaxationfac2}) into (\ref{resid3}) and rearranging will give, 
\begin{equation}\label{resid4}
  L^{i+1}  \leq  L\frac{\|S_w^{n+1,i} - S_w^{n+1,i-1}   \|}{\|\widetilde  S_w^{n+1,i+1} - S_w^{n+1,i} \|}.
\end{equation}
Recall that $L^{i+1}\in (0,1]$ and from equation (\ref{resid2}), and thus the choice of $L^{i+1}$ should satisfy instead, 
\begin{equation}\label{resid5}
  L^{i+1}  \leq  \min\Big\{L^{\rm max}, L\frac{\|S_w^{n+1,i} - S_w^{n+1,i-1}   \|}{\|\widetilde  S_w^{n+1,i+1} - S_w^{n+1,i} \|}\Big\},
\end{equation}
where $L^{\rm max}$ and $L$ are specified a \textit{priori}. In this paper, we considered $L = 0.5$ and $L^{\rm max} = 1$, see Table \ref{BtubeM}.
\section{Numerical results}\label{NumRes}
In this section, we examine the convergence and accuracy of the iterative pseudo-monolithic scheme presented in this work. Section \ref{AcaExa} presents a comparison between  the pseudo-monolithic (\ref{pressureNew})-(\ref{continutyNew}) scheme and the iterative pseudo-monolithic (\ref{pressureNew2})-(\ref{relaxationfa}) scheme. We also carry out  comparisons between iterative IMPES, and iterative pseudo-monolithic  scheme in section \ref{pysTest}. All the schemes are implemented in the open source software package MRST \cite{Knut2016}. Here, we applied two point flux approximation (TPFA) to discretize the models designed below. However, we recall that we applied a Galerkin finite elements  to show the convergence of the scheme theoretically. This is to show that the scheme is independent of space discretization methods. 
\subsection{Analytic example}\label{AcaExa}
In this subsection, a porous medium flow  model is designed by choosing exact solutions 
\begin{eqnarray}
S_w^{\rm an} = 0.65-tx(1-x)y(1-y),~  
P_o^{\rm an} = tx(1-x)y(1-y) + 0.2~~~  {\rm in}~ (0,T_f)\times \mathrm{\Omega},\nonumber
\end{eqnarray}
followed by constructing source terms and boundary conditions. For this particular example, we set $t \in [0,1]$ and $\mathrm{\Omega} = (0,1)\times  (0,1)$. Further,  we consider unit magnitude for rock as well as fluid
properties  in order to ease the construction of the source terms. We applied van Genuchten relative permeability relations (\ref{caplam}) and dynamic capillary pressure model (\ref{eq:dynPc_interp}) with $$\omega = \frac{\beta_1\chi}{\beta_1\chi + 1}.$$
The constitutive model parameters are listed in Table \ref{BtubeM}.
\begin{table}[!ht]
\centering
\begin{tabular}{l l l l }
\toprule
parameters  & values   & parameters       & values  \\
\midrule
$n_w, n_o$             &  2 &   $\beta_1$ & 400   \\
$P_e^w$                &  1 &   $P_e^o$   &  0  \\
 \bottomrule
\end{tabular}
\caption{Parameter values of relative permeabilities and capillary pressure models.}\label{BtubeM}
\end{table}

To evaluate the convergence of the scheme to the exact solution, we have considered a $80\times 80$ regular grid cells with varying time step sizes given below
\begin{equation}
\delta t\in \Big\{1/5, ~1/20,~ 1/40, ~1/60, ~    1/100, ~1/140\Big\}.\nonumber
\end{equation}
The outer iteration loop for the iterative pseudo-monolithic scheme is allowed to continue  until $\Vert S_w^{n+1,i+1} - S_w^{n+1,i}\Vert \leq 1\times 10^{-6}$ is satisfied. In this test, the  relaxation factor choice strategy mentioned in subsection \ref{relaxation} is applied. Initially, $L^1$ is computed from \eqref{resid}, where we take $\big\Vert S_w^{n+1,0}-S_w^{n+1,-1}\big\Vert$ = 1 in each time step.

\begin{figure}[h!]
\centering
\includegraphics[scale=0.5]{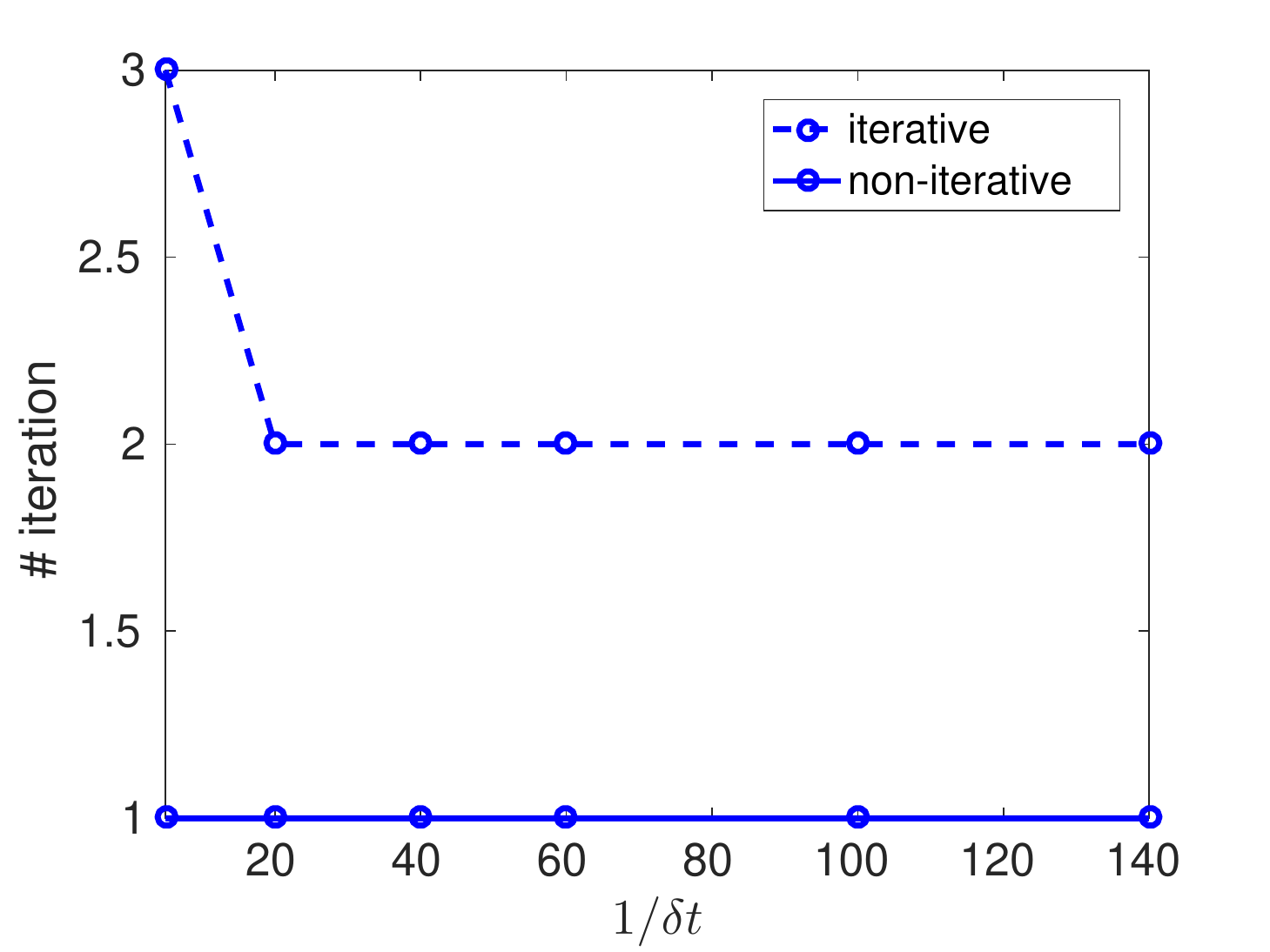}
\includegraphics[scale=0.5]{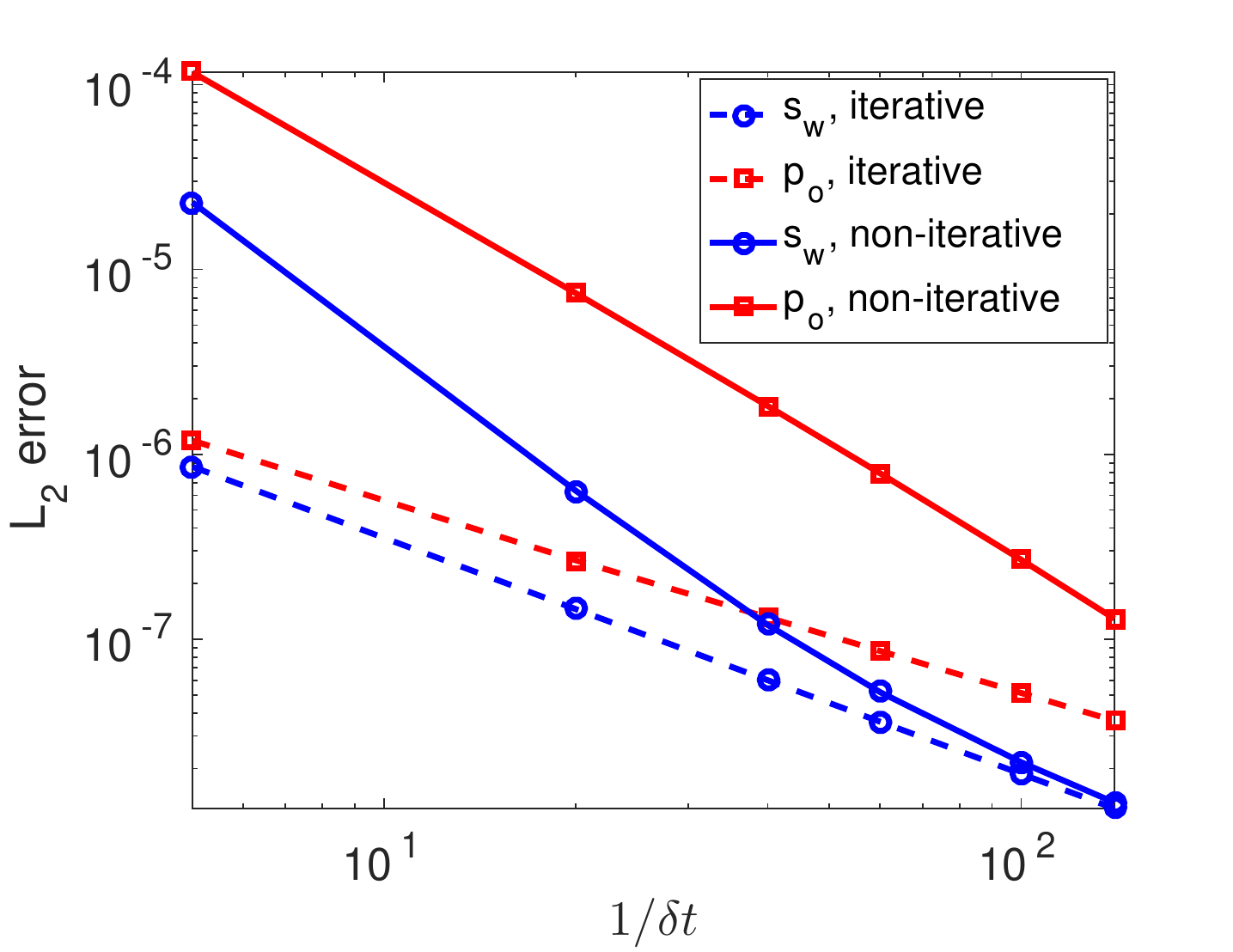}
\caption{Number of iterations (a) and $L_2$ error (b) with respect to time step sizes. }\label{lerror}
\end{figure}

We experimented a convergence test considering the inputs above, and  
Figure \ref{lerror}a presents the number of iterations of the iterative  pseudo-monolithic scheme for different time step sizes. Further, we also plotted  the number of iterations of the pseudo-monolithic scheme just as a reference. 
Obviously, the proposed pseudo-monolithic scheme exits the iteration steps at the first iteration for all time steps. On the other hand, the proposed iterative pseudo-monolithic scheme converges to the solution within two iterations for all time steps except for the larger time step $\delta t=0.2$ which needs one extra iteration.

Figure \ref{lerror}b shows the associated $L_2$ error $\Vert S_w^{n+1}-S^{\rm an}_w\Vert_{L_2(\mathrm{\Omega})}$ for the saturation and pressure profiles. $S^{\rm an}_w$ represents the analytical solution of the saturation. The pseudo-monolithic  method approximates the exact solution efficiently. The iterative pseudo-monolithic scheme has improved the accuracy of the pseudo-monolithic scheme as proposed in section \ref{Descretization}. This explains that the efficiency and accuracy of the iterative scheme can be gained with only the cost of few extra iterations. Note that the number of iterations can be reduced by considering larger stopping criteria for outer iterations without affecting the accuracy.  
 
We have also performed a numerical experiment to analyze the convergence of the proposed iterative method by fixing the time step size $\delta t$ for different number of grid cells with
\begin{equation}
h = \Big\{1/10,~1/20,~1/40,~1/60,~1/80,~1/100\Big\},
\end{equation}
where $h$ is the side length of a uniform grid cell.
The obtained results   are listed in Table \ref{lerror_2}.
\begin{table}[h!]
\centering
\begin{tabular}{l l l l  l l l}
\toprule
1/h & 10 & 20 & 40 & 60 & 80 & 100\\
\midrule 
Number of iterations($\delta t = 0.2$) & 3 & 3&3&3&3 &3\\
Number of iterations($\delta t = 0.05$) & 2 & 2&2&2&2&2\\
\bottomrule
\end{tabular}\caption{The required number of iteration to converge to the solution per time step for different mesh size resolutions. }\label{lerror_2}
\end{table}

The iterative pseudo-monolithic method converges with a maximum iteration of three. This maximum number of iteration was needed for the largest time step size $\delta t=0.2$. From Table \ref{lerror_2}, we observe that the number of iterations keeps the same while  the grid size varies. This implies that the proposed iterative pseudo-monolithic scheme is not dependent on the mesh size.

\subsection{Physical Test}\label{pysTest}
Above, we considered an academic example and  studied the accuracy and efficiency  of the iterative pseudo-monolithic method over the pseudo-monolithic method. In the following, we will compare the iterative pseudo-monolithic scheme and IMPES by considering complex porous media geometries with realistic material properties. These properties are given in Table \ref{BtubeM}.
\begin{table}[!ht]
\centering
\begin{tabular}{l l l l}
\toprule
parameters       & units        & Example 1   & Example 2 \\
\midrule
$\phi$           &     [-]      & 0.2         & 0.2         \\
$\mu_{w}$        &     [cP]     & 1           & 1            \\
$\mu_{o}$        &     [cP]     & 0.45        & 0.45          \\
$n_w$            &     [-]      & 2           & 2              \\
$n_o$            &     [-]      & 2           & 2               \\
$P_e^w$          &     [bar]    & 5           & 5               \\
$P_e^o$          &     [bar]    & 0           & 0                 \\
$L$              &     [-]      & 0.5         & 0.5                \\
$L^{\rm max}$    &     [-]      & 1           & 1                   \\
$\beta_1$        &     [-]      & 100         & -                    \\
$\beta_2$        &     [-]      &  -          & 100                  \\
\bottomrule
\end{tabular} 
\caption{Material properties and model parameters.}\label{BtubeM}
\end{table}
Note that fluid properties and model parameters given in Table \ref{BtubeM} are 
applied for both  iterative IMPES and iterative pseudo-monolithic schemes. The outer iteration loop  is allowed to continue  until $\Vert S_w^{n+1,i+1} - S_w^{n+1,i}\Vert \leq  2.5\times 10^{-5}$ is satisfied. The relaxation factor choice strategy starts with computing $L^1$  from \eqref{resid}, where we take $\big\Vert S_w^{n+1,0}-S_w^{n+1,-1}\big\Vert = 1$ in each time step. The relative permeabilities and capillary pressure models in example 1 and 2 below  are considering a zero residual saturations for the wetting and non-wetting fluids.
\subsubsection{Example 1}
\begin{figure}[h!]
 \centering
 \includegraphics[width=3in, height = 2in]{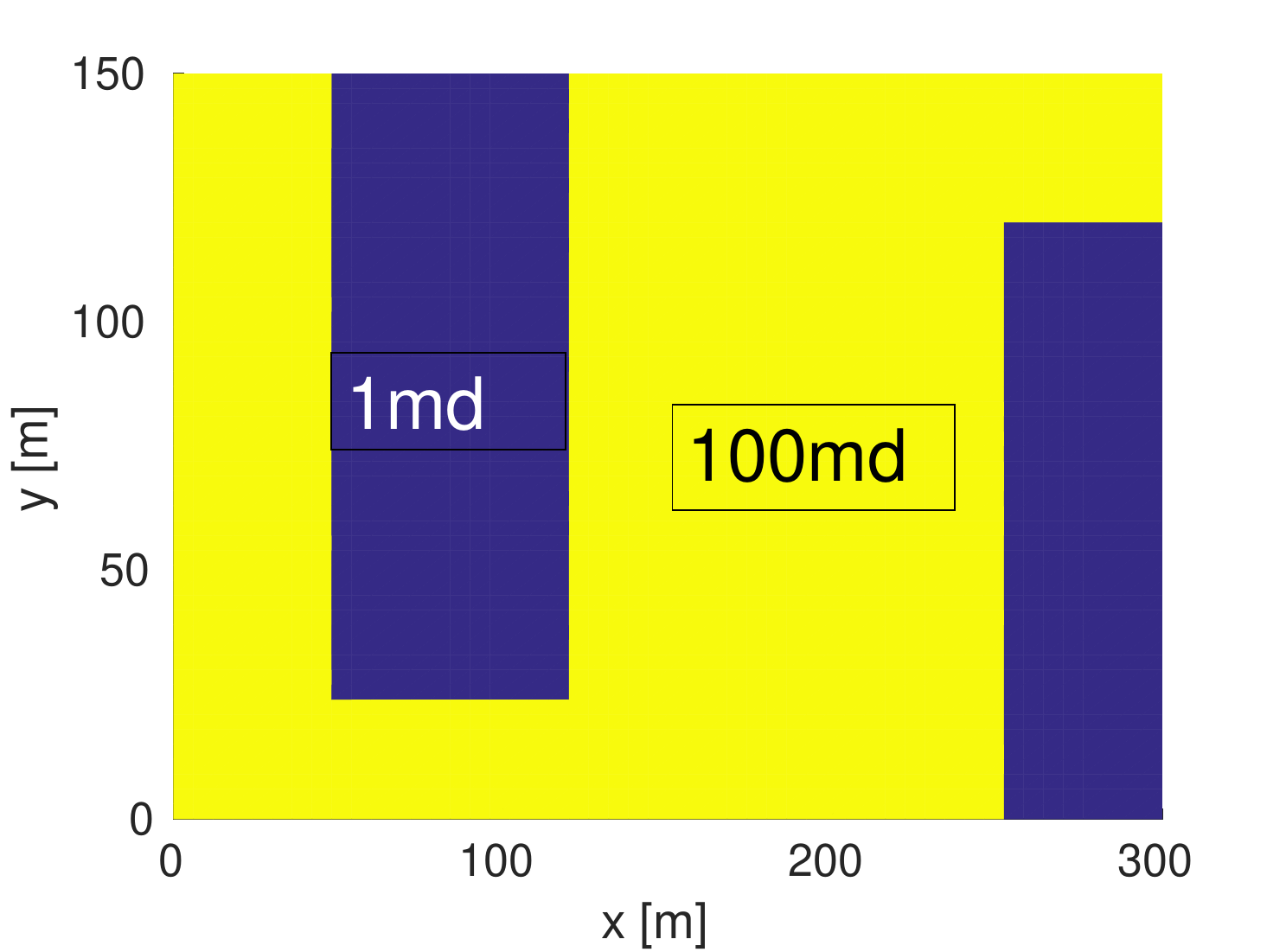}
 \caption{Rock permeability model for example 1. Here "md" stands for milli darcy.}\label{reckPerm}
 \end{figure}

The computational domain with  $300 {\rm m}\times 150 {\rm m}$ dimensions,
 consisting  of different sub-domains  for the distribution of permeability, is considered. This porous medium model is shown in Figure \ref{reckPerm}.
 
We applied the water-wet van Genuchten relative permeabilities (\ref{caplam})  and a capillary pressure function as given below,
\begin{equation}\label{dyc_pc}
P_c = \frac{\beta_1\chi}{1 + \beta_1\chi}\Big(P_c^{ow}-P_c^{ww}\Big) + P_c^{ww},
\end{equation}
where $P_c^{ww}$ and $P_c^{ow}$ are as described in Equation (\ref{cap-p}).
The capillary pressure (\ref{dyc_pc}) is changing from $P_c^{ww}$ to $P_c^{ow}$ dynamically for 7.5 years. Here 7.5 years represent the life of injection for this particular simulation. 
Further data on the  model parameters   are given in Table \ref{BtubeM} above. We complete the model by injecting the non-wetting fluid  to the left-bottom corner of the domain with an injection rate of $0.35{\rm m^3}$ per day and we impose a zero Dirichlet condition at the middle of the right side of the domain. The rest of the boundaries are considered impermeable. 

We  discretized the above model with 2500 regular grid cells, and performed numerical experiments to evaluate the convergence behavior of the iterative IMPES and pseudo-monolithic scheme for different time step sizes. Table \ref{BtubeM_comp} shows the convergence results of the two methods.  
\begin{table}[!ht]
\centering
\begin{tabular}{l l l l l l l l  }
\toprule

$\delta t$ (days)& 27.5   & 6.85  & 3.43   & 1.71 & 0.85 \\ 
\midrule
\underline{ iterative IMPES} & & &&\\
Total iteration      & -      & -  &  -  & -   & 12237 \\
Average iteration    & -      & -  &  -   & -  & 3.82\\
\underline{iterative pseudo-monolithic}    & & & &\\

Total iteration      & 672     & 1820  & 2832  & 4112  & 5244 \\
Average iteration    & 6.72    & 4.55  & 3.45  & 2.57  & 1.6\\
\bottomrule
\end{tabular} 
\caption{Linearization schemes convergence comparision for the tested flow model in example 1. The '-' sign stands for the scheme is not convergent, ''Total iteration'' stands for the over all number of iterations to complete the simulation, and ''Average iteration'' represents the average iteration number per  time step. }\label{BtubeM_comp}
\end{table}
As shown in Table \ref{BtubeM_comp},   the iterative IMPES only converges if the time step size $\delta t \le 0.85$ day, and the iterative pseudo-monolithic scheme converges for all time step sizes. This shows that the iterative IMPES is subject to strong restrictions with respect to  the time step size choice. Usually, IMPES encountered a difficulty regarding the choice of time step size even for standard multi-phase flow models  \cite{Kou2010, Coats2003,  Birane2019, Kou2010b}. The dynamic nature of the capillary pressure function further worsens the flexibility of iterative IMPES  on the choice of the time step size in this example.  In contrast, the iterative pseudo-monolithic scheme shows its strength allowing for relaxed choice of time step size. The scheme is capable of taking large time step size. Furthermore, the total and average number of iterations are respectively increasing and decreasing  while the scheme considers smaller time step sizes. Decreasing number of average iteration per time step size is  a positive sign towards the stability of the iterative pseudo-monolithic scheme.

We further studied the convergence stability of the iterative pseudo-monolithic scheme by controlling the speed of capillary pressure alteration. To do so,  we vary the dynamic coefficient parameter $\beta_1$ from Equation (\ref{dyc_pc}). For this numerical experiment, we used the same mesh size as before and chose larger time step size $\delta t = 30.5$ days. Table \ref{sta_ab_1} shows the convergence behavior of the iterative pseudo-monolithic scheme for different  dynamic coefficient parameter, $\beta_1$, values.
\begin{table}[!ht]
\centering
\begin{tabular}{l l l l l l}
\toprule
~~~~~$\beta_1$   & 100 & $200$ &  $10^{4}$ & $2\times 10^{4}$     \\
\midrule
Total iteration      & 622    & 644   & 768    & 963        \\
Average iteration      & 6.9  & 7.2   & 8.5    & 10.7     \\
\bottomrule
\end{tabular}
\caption{The impact of dynamic coefficient parameter on the stability of the iterative pseudo-monolithic scheme.}\label{sta_ab_1}
\end{table} 
From Table \ref{sta_ab_1}, we observe that the scheme requires more iterations as $\beta_1$ increases. That means the scheme needs a few extra iterations  to converge as the alteration speed of the capillarity becomes faster. Further, the scheme may fail to converge for this model if we choose sufficiently large $\beta_1$ (not shown here). For such case, the proposed scheme is enforced to choose relatively larger time step size. 
Nevertheless, the results above show that the scheme converges successfully for non-local (in time) two-phase flow model that considers physically reasonable dynamic capillary pressure alteration (even with capillary pressure jumps). These all support  the theoretical convergence analysis of the proposed scheme, where discussed in section \ref{Descretization}. In general, the reliability of the scheme to handle the dynamic alteration of the capillary pressure and non-locality of the problem has been successfully demonstrated.

\begin{figure} 
\centering 
\includegraphics[scale=0.5]{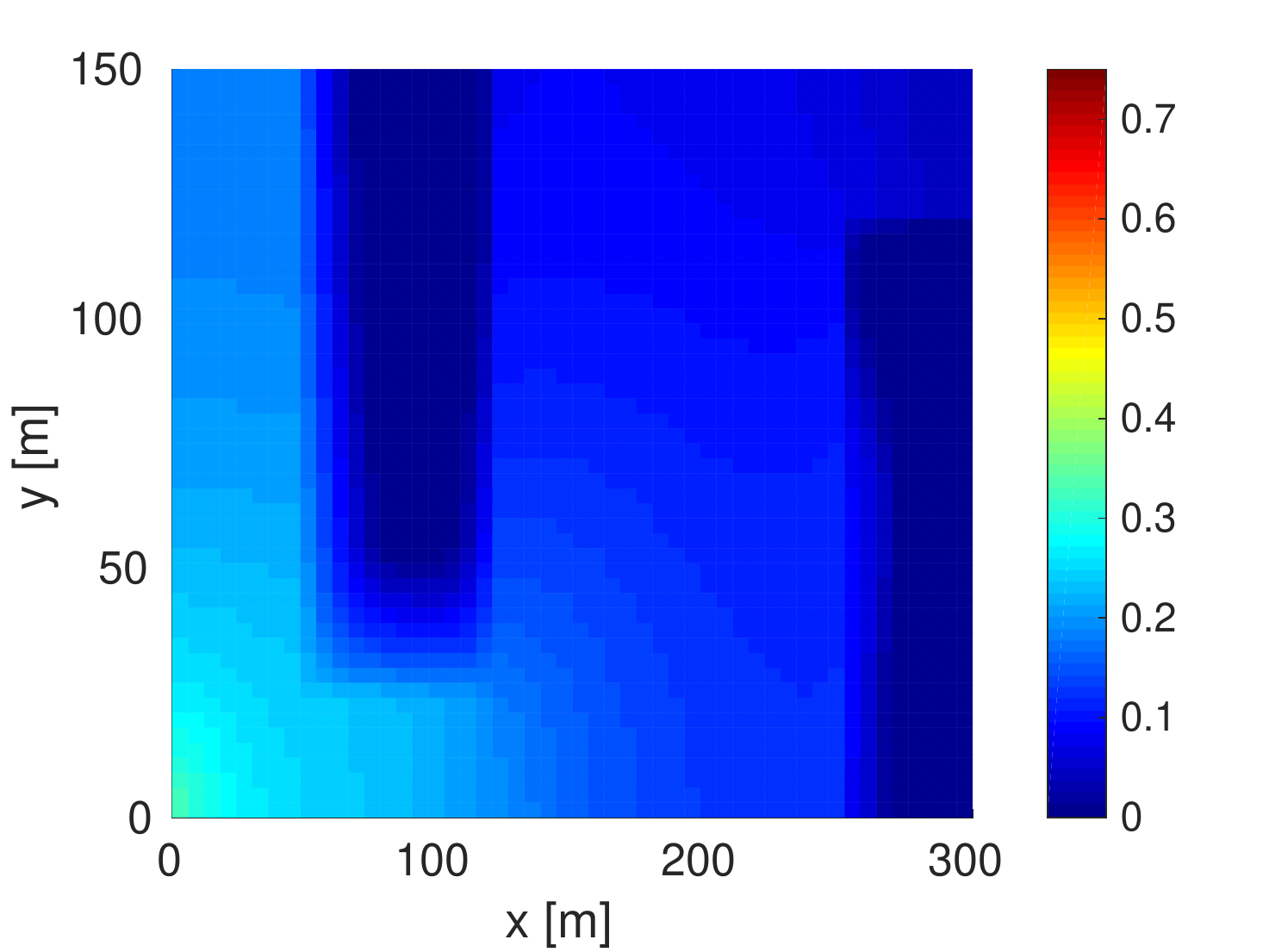}
\includegraphics[scale=0.5]{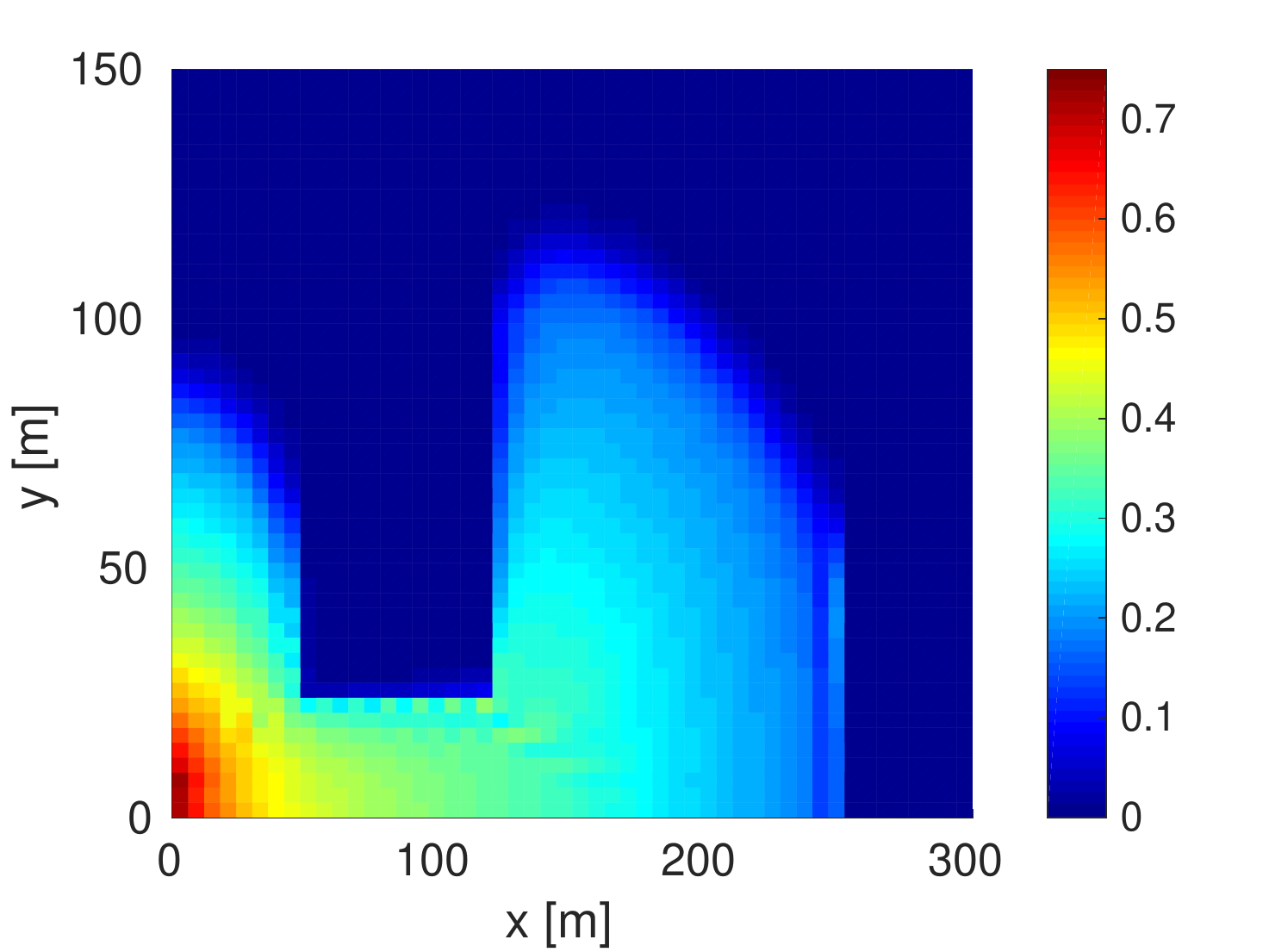}
\caption{Saturation profiles: (a) demonstarates the flow path when $\beta_1 = 1$, i.e., slow dynamic capillary pressure alteration and (b) shows the fluid path for relatively fast dynamic alteration, i.e., $\beta_1 = 2\times 10^{4} $.}\label{fig_real_exa_1}
\end{figure}

After a successful convergence stability experiment, we also studied the impact of the dynamic coefficient parameter, $\beta_1$, on the flow path of the fluids. We used  $\delta t = 30.5$ days, and we keep the   grid size,  fluid and reservoir parameters as above for this purpose. However, we considered two different values for the dynamic coefficient parameter $\beta_1$. Figure \ref{fig_real_exa_1} shows the fluid distributions after 7.5 years of evolution. 
As shown in Figure \ref{fig_real_exa_1}, the effect of the dynamic coefficient on fluid displacement is distinct. The movement of the non-wetting fluid is restricted for fast dynamic alteration processes. In other words, the displacing fluid remains the resident fluid for slow dynamic capillary pressure alteration, whereas it swipes the resident fluid when we consider a fast capillary pressure alteration. This happens because  the wetting property of the volumes (occupied by the displacing fluid) are altered to intermediate-wet system before it leaves the volume, and thus, the non-wetting fluid preferred to be in contact with the solids when we consider fast capillary pressure alteration.

\subsubsection{Example 2}
We considered  $50{\rm m}\times 50{\rm m}\times 10 {\rm m}$ dimensional heterogeneous medium, with permeability distribution,
\begin{eqnarray}
\mathtt{K} = \left\{\begin{array}{l}
 1~ {\rm md}, ~{\rm if}~ (x,y,z) \in (5{\rm m}, 50{\rm m})\times (0{\rm m}, 30{\rm m})\times (2{\rm m}, 8{\rm m}) \\[0.06in]
100~ {\rm md}, {\rm else}.
\end{array}\right.
\end{eqnarray}
We employed the same relative permeabilities functions as above  and dynamic capillary pressure given as,
\begin{equation}\label{dyc_pc_2}
P_c = \frac{\beta_2S_w\chi}{1 + \beta_2S_w\chi}\Big(P_c^{ow}-P_c^{ww}\Big) + P_c^{ww},
\end{equation}
where $P_c^{ww}$ and $P_c^{ow}$ are as described in Equation (\ref{cap-p}).
The capillary pressure is allowed to change from $P_c^{ww}$ to $P_c^{ow}$ dynamically according to model (\ref{dyc_pc_2}) in each sub-domain  for 2.5 years. Additional data on model parameters are listed in Table \ref{BtubeM}. We inject the non-wetting fluid to the west particularly at $(y,z) \in  (10{\rm m}, 15{\rm m})\times (2{\rm m}, 8{\rm m})$ with an injection rate of $0.15{\rm m^3/day}$ for 2.5 years, and  impose a zero Dirichlet  boundary condition to the east side of the domain, particularly at $(y,z) \in  (10{\rm m}, 15{\rm m})\times (2{\rm m}, 8{\rm m})$. The rest of the boundaries are considered to be impermeable. 

We discretized the designed model above with 3125 grid cells and we did  simulations to examine the convergence behavior of the iterative IMPES and pseudo-monolithic linearization techniques. The obtained results are published in Table \ref{BtubeM_comp_5}.
\begin{table}[h!]
\centering
\begin{tabular}{l l l l l l l l }
\toprule
$\delta t$(days)           &90.25  & 60   & 30.4       & 9.125   &  2.2812 & 0.57 & 0.26\\
\midrule
\underline{iterative IMPES}&  &           &     &     &       &   &              \\
Total iteration            &-  & -         & -   & -   & -    & -  & 6980       \\
Average iteration           &- & -         & -   & -   & -    & -  & 2.2        \\
\underline{iterative pseudo-monolithic} &      &     &    &      \\
Total iteration       &50 & 72     & 121    &   271     &  824   &  2106   &   3354    \\
Average iteration     &5 & 4.8   & 4.03   &   2.71    &  2.06   & 1.3    &  1.05 \\
\bottomrule
\end{tabular} 
\caption{Linearization schemes comparision for the tested flow model above in example 2. Here "-" represents that the scheme is not convergent.}\label{BtubeM_comp_5}
\end{table}
In Table \ref{BtubeM_comp_5}, we noticed that the iterative IMPES fails to converge for time step sizes bigger than 0.26 day. This shows that the choice of a time step size is strongly restricted for iterative IMPES linearization which confirms the results reported in \cite{Kou2010,Birane2019,Kou2010b}. Unlike the iterative IMPES, iterative pseudo-monolithic scheme relaxes the choice of the time step size. The scheme return with an approximate solution for relatively large time step size compared to the iterative IMPES scheme. 
The total and average number of iterations respectively are    increasing and decreasing when we consider smaller time step sizes, see Table \ref{BtubeM_comp_5}.  

We further investigate the sensitivity of the dynamic coefficient parameter $\beta_2$. For this, we keep the number of grid elements as before and the time step size to be the larger one in Table \ref{BtubeM_comp_5}, i.e., $\delta t = 90.25$ days. Then, we  vary    $\beta_2$, and observe its impact on the convergence  of the  scheme. Table \ref{sta_ab} shows the convergence results for different  values of $\beta_2$.
\begin{table}[h!]
\centering
\begin{tabular}{l l l l l l l l l}
\toprule
~~~~~$\beta_2$       & 100      & 1000 & $2\times 10^3$ & $1\times 10^{4}$& $2\times 10^{4}$ \\
\midrule
Total iteration      & 50       & 50   &  50   & 50   & 50 \\
Average iteration    & 5        & 5    &  5  & 5  & 5 \\
\bottomrule
\end{tabular}
\caption{The impact of dynamic coefficient parameter on the stability of the iterative pseudo-monolithic scheme.}\label{sta_ab}
\end{table}
As shown in Table \ref{sta_ab}, the scheme converges with the same number of iterations for all values of $\beta_2$. This implies that the proposed scheme is not affected by the the speed of the capillary pressure alteration dynamics.

\begin{figure}[h!]
\centering 
\includegraphics[scale=0.5]{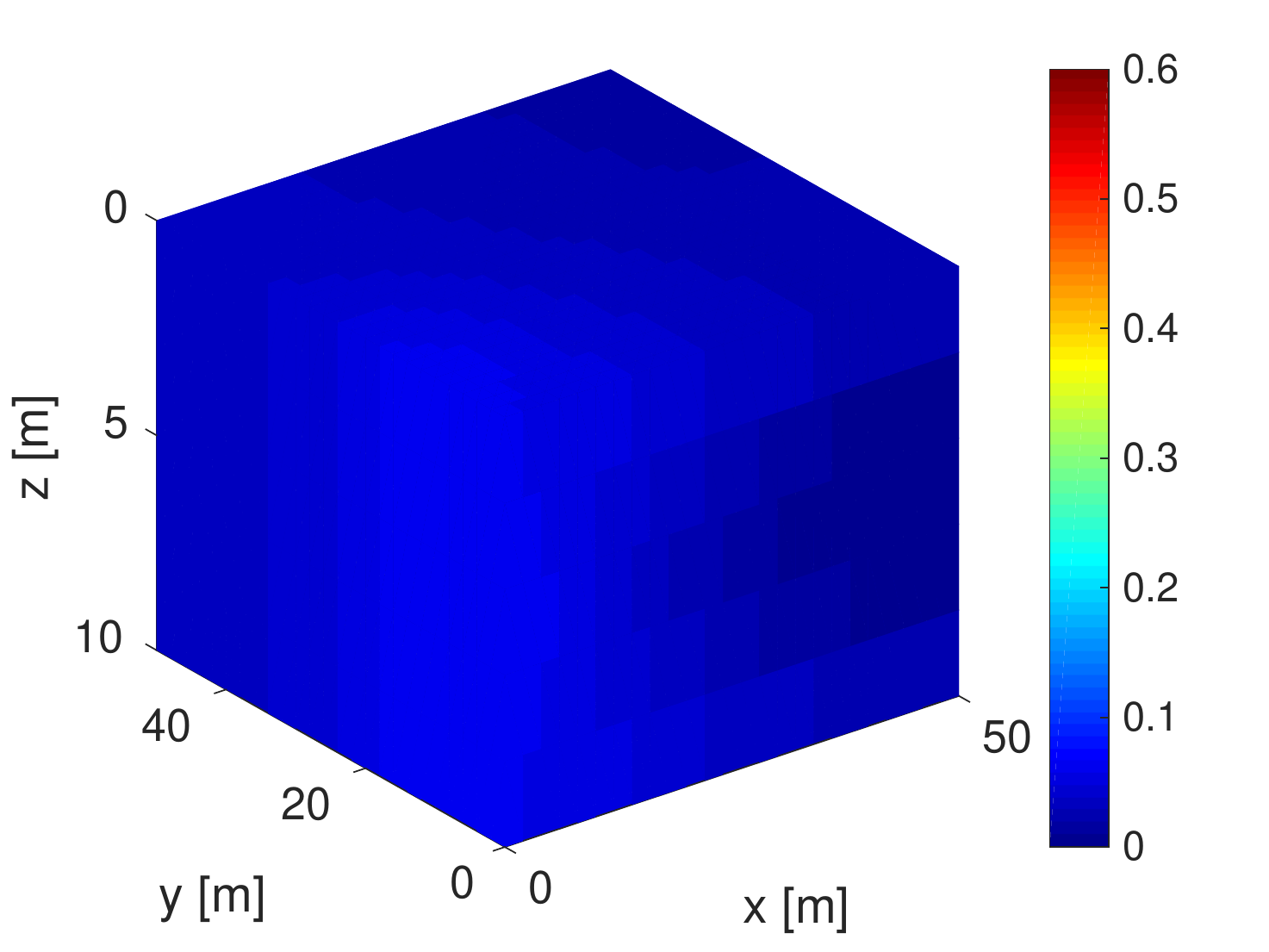}\includegraphics[scale=0.5]{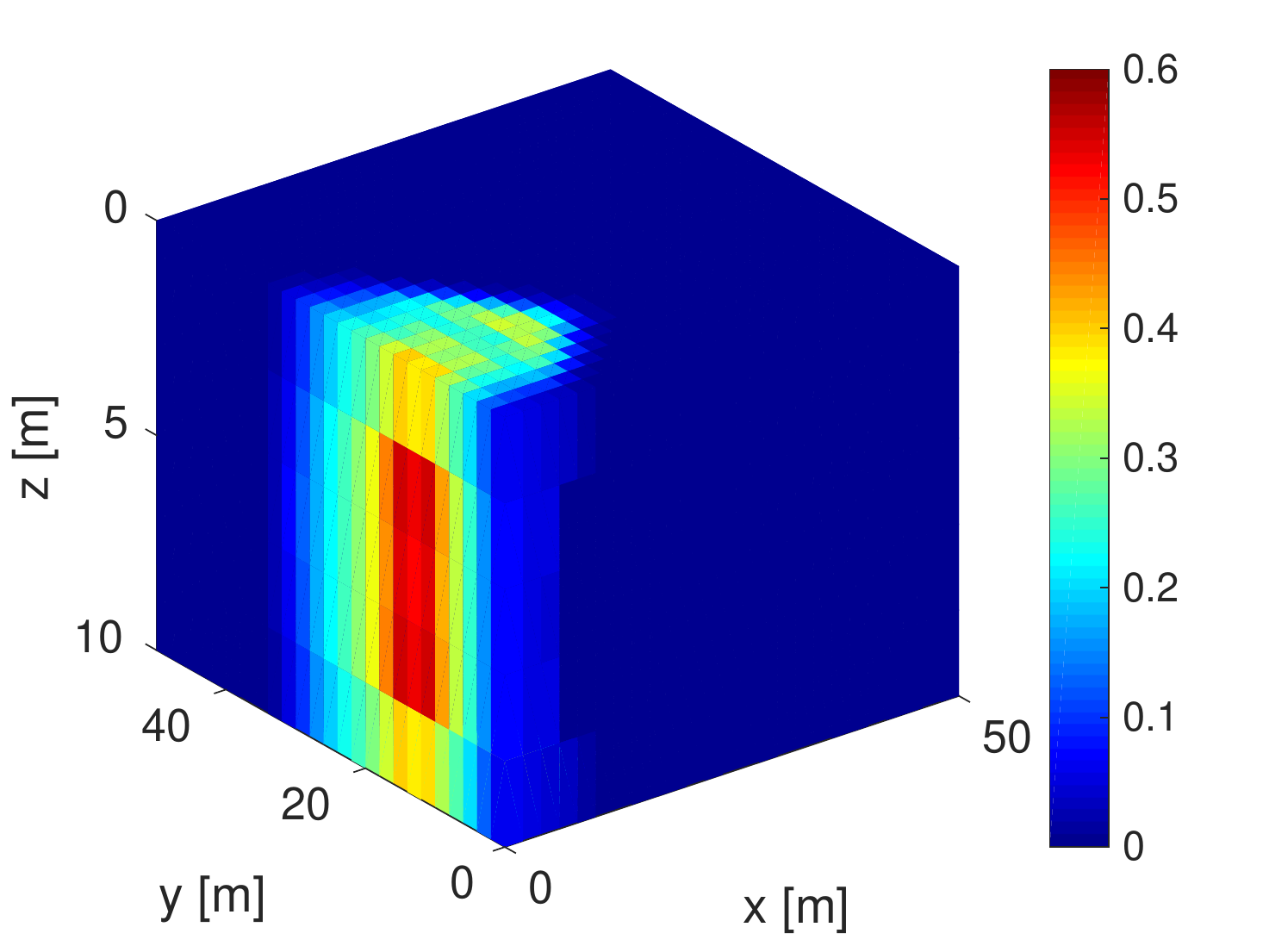}
\caption{Saturation distribution obtained from iterative pseudo-monolitich scheme. Here, we applyied the same grid resolution as above with time step size $\delta t = 90.25$ days. (a) no dynamic capillary pressure alteration ($\beta_2=0$) and (b) relatively fast alteraion ($\beta_2=1\times 10^5$ in Equation (\ref{dyc_pc_2})).}\label{fig_real_exa_2}
\end{figure}

Above, we studied the convergence of the iterative pseudo-monolithic scheme for non-local two-phase flow model. Below, we investigate the impact of the dynamic capillary pressure model on the injected fluid distribution. Figure \ref{fig_real_exa_2} compares the injected fluid distribution for the initial wetting condition capillary pressure (i.e., $\beta_2=0$ in Equation (\ref{dyc_pc_2})), and dynamic capillary pressure model (\ref{dyc_pc_2}) with $\beta_2 = 1\times 10^5$.
In Figure \ref{fig_real_exa_2}, we observe that the displacing fluid leaves the resident fluid behind when we consider $\beta_2 =0$ in Equation (\ref{dyc_pc_2}). This is due to the fact that the rock surfaces are water-wet in this case, i.e., no WA, and thus, the resident fluid prefers to remain in the pores. 
On the other hand, the non-wetting fluid displaced the resident fluid and concentrated near the injection area when we employed the dynamic capillary pressure model (\ref{dyc_pc_2}) with $\beta_2 =  1\times 10^5$. In this case, the wettability of the rock surfaces near to the injection area have been changed (in time) to intermediate-wet system before the displacing fluid leaves the volume, and thus, the displacing fluid preferred to occupy these pores. That means, the injected fluid pressure are able to displace the resident fluid with relatively small pressure. This shows that the dynamic capillary pressure results in a large change of fluid saturation as compared to the standard capillary pressure model, $P_c^{ww}$, in Equation (\ref{cap-p}). This might be one of the reasons that restricts the time step size choice of the iterative IMPES.

\section{Conclusion}\label{Con}
In this paper, we introduced fluid history and time-dependent dynamic capillary pressure model in a two-phase immiscible incompressible porous media flow model. 
We developed a linearization scheme for the resulting non-standard two-phase flow model by treating the capillary pressure  implicitly and adding stabilization terms. This implicit treatment of  the dynamic capillary pressure model couples the pressure and saturation equations strongly, and makes the scheme stable. We gave a theoretical convergence analysis of the scheme under some meaningful assumptions. The scheme has been successfully implemented and tested for different illustrative examples. We found that the proposed scheme is efficient to approximate the solution of the resulting non-standard two-phase flow model. Most importantly, the scheme demonstrates flexibility regarding the choice of time step size for dynamic  capillary pressure alteration (possibly with capillary jumps). Thus, combining the scheme with a Newton method is a straightforward application. This implies that one can alternate between the iterative pseudo-monolithic scheme and Newton method as mentioned in \cite{List2016}. This may further improve the convergence speed and accuracy of the approximation to simulate such complex models. .

\bibliographystyle{numeric}

\end{document}